\documentclass[twocoloumn,prb,amssymb,reprint,superscriptaddress,showpacs 
 ]{revtex4-1}

\usepackage{hyperref}
\usepackage{bm}
\usepackage[dvips]{graphicx}
\usepackage{amsmath}
\usepackage{natbib}

\begin{document}

\title{Quantum confinement, energy spectra and backscattering of Dirac fermions in quantum wire in magnetic field}%

\author{V.V. Enaldiev}
\email[e-mail:]{vova.enaldiev@gmail.com}
\affiliation{V.A. Kotelnikov Institute of Radio-engineering and Electronics of Russian Academy of Sciences, Mokhovaya st. 11-7, Moscow, 125009 Russia}
\affiliation{Moscow Institute of Physics and Technology, 141700, Institutskii per. 9, Dolgoprudny, Moscow Region, Russia\\}
\author{V.A. Volkov}
\email[e-mail: ]{volkov.v.a@gmail.com}
\affiliation{V.A. Kotelnikov Institute of Radio-engineering and Electronics of Russian Academy of Sciences, Mokhovaya st. 11-7, Moscow, 125009 Russia}
\affiliation{Moscow Institute of Physics and Technology, 141700, Institutskii per. 9, Dolgoprudny, Moscow Region, Russia\\}

\date{\today}

\begin{abstract}
We address the problems of an energy spectrum and backscattering of massive Dirac fermions confined in a cylindrical quantum wire. The Dirac fermions are described by the 3D Dirac equation supplemented by time-reversal-invariant boundary conditions at a surface of the wire. Even in zero magnetic field, spectra quantum-confined and surface states substantially depend on a boundary parameter $a_0$. At the wire surface with $a_0>0$ ($a_0<0$) the surface states form 1D massive subbands inside (outside) the bulk gap. The longitudinal magnetic field transforms the energy spectra. In the limit of the thick wires and the weak magnetic fields, the 1D massless surface subbands arise at half-integer number of magnetic flux quanta passing through the wire cross section. We reveal conditions when backscattering of the surface Dirac fermions by a non-magnetic impurity is suppressed. In addition, we calculate a conductance formed by the massless surface Dirac fermions in the magnetic field in collisional and ballistic regimes. 
\end{abstract}


\maketitle

\section{Introduction}
For the last few years the concept of topological insulators has attracted great attention to a study of massless surface (or edge) Dirac fermions (DFs) in crystals. These excitations can be realized in semiconductor structures of two types: 1) at a flat surface of the 3D topological insulators (TIs), like $Bi_2Se_3$, $Bi_2Te_3$\cite{Zhang_Liu,Liu_Zhang} or at an edge of the 2D TIs in $CdHgTe$ quantum wells; 2) at a flat surface of narrow-gap semiconductors\cite{volkov_pinsker}, like bismuth, bismuth-based solid alloys, lead chalcogenides. In a continuous model both systems are described by multicomponent envelope functions. However, it is assumed that the surface states (SSs) in the two types of  systems arise due to different reasons. Emergence of the SSs in the TIs is caused by changing of the topological $Z_2$ indices across a crystal--vacuum interface\cite{Kane_Hasan,Qi_Zhang}. The topological indices are solely defined by the Bloch Hamiltonian\cite{Fu_Kane_Mele}. Because they are robust to moderate perturbations of the Bloch Hamiltonian and can change only by closing of the gap, the SSs are topologically protected. For the systems of the second type, the SSs arise due to an abrupt cutoff of a crystal potential, as in Refs.[\onlinecite{Tamm_IE},\onlinecite{Shokley_W}]. In latter case the topological arguments are not relevant. These SSs are sometimes called the Tamm states or the Shockley states, or the Tamm--Schockley states.  

3D TIs ($Bi_2Se_3$-type) are described in terms of the envelope functions by a modified 3D Dirac Hamiltonian with momentum depending mass term\cite{Zhang_Liu,Liu_Zhang,Silverstov_brouwer} $m(\bm{p})~=~M-~M_1\bm{p}^2$:
\begin{equation}\label{TI_ham}
		H_{TI}= \left ( 
				\begin{array}{cc}
				m(\bm{p})\sigma_0 & A\bm{\sigma p} \\
				A\bm{\sigma p} & -m(\bm{p})\sigma_0 
				\end{array}
						\right ),
\end{equation}
where $\bm{p}=(p_x,p_y,p_z)$ is the 3D momentum operator, $A,M,M_1$ are the model parameters, $\bm{\sigma}=(\sigma_x,\sigma_y,\sigma_z) $ is the vector of the Pauli matrices, $\sigma_0$ is an identity matrix 2$\times 2$. To derive the surface state spectrum of $H_{TI}$, one needs to specify boundary conditions (BCs) for the envelope functions. The Hamiltonian $H_{TI}$ has the second order in the momentum operators, therefore it is required four restrictions for the envelope functions at the surface. Since the envelope functions are only a smoothly varying factor of a real wave function, the problem of the proper BCs for them is not trivial. One of the most widespread and the simplest BCs are the open BCs\cite{Zhang_Liu,Linder_J,LuHZ,Shan_W}. These BCs guarantee appearing of the SSs in the bulk gap and agree with a spectrum of the tight--binding calculations for the (111) surface\cite{Pershoguba_Yakovenko}. However, the issue of correct BCs for other surface termination is poorly studied. 

In the systems of the second type, the envelope functions of the massive Dirac fermions obey the 3D Dirac equation with a constant mass term\cite{Keldysh_LV,Wolff_PA}:
\begin{equation}\label{Dirac_hamiltonian}
		 \left ( 
				\begin{array}{cc}
				mc^{2}\sigma_0 & c\bm{\sigma p} \\
				c\bm{\sigma p} & -mc^{2}\sigma_0 
				\end{array}
						\right )
\left ( \begin{array}{c}
\Psi_c\\
\Psi_v
\end{array}
\right )=
E\left ( \begin{array}{c}
\Psi_c\\
\Psi_v
\end{array}
\right ),
\end{equation}
where $m$ is an effective mass, $c$ is an effective ''speed of light''. In eq. (\ref{Dirac_hamiltonian}) two-component spinors $\Psi_c$, $\Psi_v$ are the envelope functions corresponding to the conduction (c) and valence (v) bands. In this case we need only two restrictions for components of the spinors at the surface, as the Dirac Hamiltonian is of the first order in the momentum operators. One can derive the BC for the spinors from the Hermiticity and time-reversal symmetry of the Dirac Hamiltonian in a region with the surface $S$\cite{volkov_pinsker}: 
\begin{equation}\label{boundary_condition_main}
		\left ( \sigma_0\Psi_v-ia_0\bm{\sigma n}\Psi_c \right )_{\bm{r}\in S} = 0,
\end{equation}
where $\bm{n}=\bm{n}(S)$ is an inner normal to the surface $S$, $a_0$ is the real phenomenological parameter, which characterizes both the microscopic surface structure and bulk band structure. 

\begin{figure*}
  \includegraphics[width=16cm]{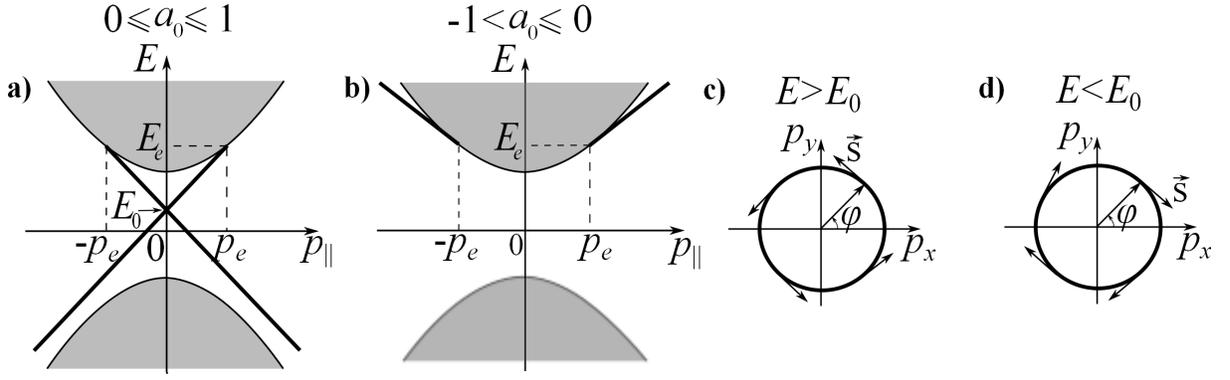}
			\caption{ \label{Fig:Semispace_specrtum} 
 Energy spectrum\cite{volkov_pinsker} of the Tamm--Dirac states (bold solid lines) of the 3D Dirac equation (\ref{Dirac_hamiltonian}) in a halfspace $z\geq 0$ with the BCs (\ref{boundary_condition_main}) for two classes of surfaces; $\bm{p}_{||}=(p_x,p_y,0)$ is an in-plane momentum. a) For the surface with the boundary parameter $0\geq a_0\leq 1$, the TD states lie inside of the bulk gap, on a cone with the Dirac point $E_0$ and end points at $p_e=2|a_0|mc/(1-a_0^2)$; b) For the surface with $-1<a_0<0$ the TD states lie on a part of the cone outside the gap; c)-d) Spin polarization $\vec{\bm{s}}$ of the TD states at the cone surface, corresponding to $s=\pm1$ and $0<a_0\leq 1$. Grey color corresponds to the bulk states continuum. }
\end{figure*}

In a halfspace $z\geq 0$ the 3D Dirac equation (\ref{Dirac_hamiltonian}) with the BCs (\ref{boundary_condition_main}) results in appearing of the 2D massless SSs with a conical spectrum\cite{volkov_pinsker}(see Fig.\ref{Fig:Semispace_specrtum})
\begin{equation}\label{dirac_spectrum}
E=-sv|\bm{p}_{||}|+E_0, 
\end{equation}
where \mbox{$v=2a_0c/(1+a_0^2)$} is the SSs speed, \mbox{$\bm{p}_{||}=(p_x,p_y,0)$}, \mbox{$E_0=mc^{2}(1-a_0^2)/(1+a_0^2)$} is the energy of the Dirac point counted from the middle of the bulk gap, here and below an effective chirality $s=\pm 1$ are eigenvalues of an operator $\sigma_z\otimes (\bm{\sigma},[\bm{n},\bm{p}])$. We will call these states the Tamm--Dirac (TD) states. One can classify the surface properties depending on a sign of the boundary parameter $a_0$. At $a_0\geq 0$ the energy of the Dirac point $E_0$ is in the bulk gap, fig.\ref{Fig:Semispace_specrtum}a, that is the spectrum of the TD states is similar with the spectrum of the SSs of TI. At $a_0<0$ the TD states spectrum is away from the bulk gap, fig.\ref{Fig:Semispace_specrtum}b. For the surfaces with $a_0=1$, the spectrum of the TD states possesses the particle--antiparticle symmetry with $E_0=0$. It should be noted, that such a case is the most popular in the theory of TIs. Below, we will show that deviation from this symmetry ($a_0\neq 1, E_0\neq 0$) has important consequences for backscattering between the TD states in a quantum wire. 

One can try to clarify the physical meaning of the parameter $a_0$ in a model of sharp ''inverse  heterocontact''\cite{volkov_pankratov,Idlis_Usmanov} (see also Ref.\onlinecite{Jackiw_Rebbi}). The model is described by the 3D Dirac equation (\ref{Dirac_hamiltonian}) with abruptly varying mass and work function (it is not included in (\ref{Dirac_hamiltonian})) across the heterocontact. The surface with of $a_0>0$  corresponds to an inversion of mass sign on the heterojunction\cite{Idlis_Usmanov}. Deviation of $a_0$ from unity accounts for particle--antiparticle asymmetry of the interface.

Thus, the massless 2D SSs arise both at a flat surface of the TI with the Hamiltonian (\ref{TI_ham}) and at a flat surface of crystal, which is described by the 3D Dirac equation (\ref{Dirac_hamiltonian}) with the BCs (\ref{boundary_condition_main}). Note that $H_{TI}$ (\ref{TI_ham}) reduces to the Dirac Hamiltonian in the limit $M_1=0$. Hence, it is reasonable to correlate results for the SSs of the two types of systems. Spin polarization of the TD states is \mbox{$\langle\bm{s}_D\rangle=[sE_0/2mc^2](\sin\varphi,-\cos\varphi,0)$}, where $\bm{s}_D=\sigma_0\otimes\bm{\sigma}/2$ is the conventional spin operator for the Dirac Hamiltonian\cite{Beresteckii_Lifshitz}, $\bm{p}_{||}=|\bm{p}_{||}|(\cos\varphi,\sin\varphi,0)$. In the case of particle--antiparticle symmetry ($a_0=1$), the TD states are not spin polarized ($\langle\bm{s}_D\rangle=0$). The topological SSs at the (111) surface of $Bi_2Se_3$ has spin polarization  \mbox{$\langle\bm{s}_{TI}\rangle=(-\sin\varphi,\cos\varphi,0)s/2$}. Recently it has been shown, that the Dirac point energy and the spin polarization of the topological SSs depend on crystallographic orientation of the surface\cite{Silverstov_brouwer,Zhang_Kane_Mele}, as $Bi_2Se_3$ is an anisotropic crystal. Consequently, more appropriate BCs for $H_{TI}$ (\ref{TI_ham}), that are differed from the open BCs, can lead to additional dependence of spin polarization and the Dirac point energy on the surface orientation. 

Consider now an effect of a longitudinal magnetic field on the spectrum of the DFs, confined in a quantum wire. 
From the one hand, a diagonalization of $H_{TI}$ (\ref{TI_ham}) in a quantum wire in the magnetic field is a sophisticated problem. To the best of our knowledge, the problem is not solved. From the other hand, the BCs (\ref{boundary_condition_main}) allow to solve the 3D Dirac equation (\ref{Dirac_hamiltonian}) and to find analytically a spectrum of the surface DFs in the magnetic field, taking the quantum confinement into account. 

In the present paper we study in frames of latter approach properties of the TD states in a cylindrical quantum wire allowing two possible classes of surfaces (i.e. different signs of $a_0$).

Recently, Aharonov--Bohm magnetoresistance oscillations have been observed in nanowires of $Bi_2Se_3$ and $Bi_2Te_3$\cite{Peng_H,FXiu}. Period of the oscillations corresponded to passing of the one flux quantum $\Phi_0$ through the nanowire cross section area. It was suggested that the effect was caused by the topological SSs. Spectrum of these SSs in the quantum wire consists of 1D surface subbands\cite{Egger_Zazunov} and has periodic dependence on a magnetic flux $\Phi$ through the wire. Due to the Berry phase $\pi$, massless surface subbands emerge at half-integer values of $\Phi/\Phi_0$. In Refs.[\onlinecite{Zhang_Vishwanath},\onlinecite{Bardarson_Brouwer}] a SS conductance along a TI quantum wire was numerically calculated in two models. First, the conductance was calculated in the model of an effective surface Hamiltonian\cite{Bardarson_Brouwer} for different disorder strength and various doping regimes. At the low doping (when the Fermi level is in the vicinity of the Dirac point) the conductance is formed by the perfectly transmitted mode. It approaches $e^2/h$ at half-integer values of $\Phi/\Phi_0$ for every disorder strength, because of suppression of backscattering in this model. At the high doping the conductance oscillates with $\Phi_0$ period only at weak disorder. At strong disorder oscillations of the magnetoconductance disappear in this regime. Second, in the Fu--Kane--Mele lattice model\cite{Zhang_Vishwanath} small gap arose in 1D subband spectrum at half-integer values of $\Phi/\Phi_0$, due to time-reversal symmetry breaking in the magnetic field. This results in a deviation of the wire conductance from $e^2/h$.  

In our paper we focus on the solution of the 3D Dirac equation supplemented by the BCs (\ref{boundary_condition_main}). One of goals constitutes in a qualitative comparison between results obtained in frames of (\ref{TI_ham}) and (\ref{Dirac_hamiltonian})-(\ref{boundary_condition_main}). Refs.[\onlinecite{Egger_Zazunov},\onlinecite{Zhang_Vishwanath},\onlinecite{Bardarson_Brouwer}] present the results of the TI Hamiltonian diagonalization.  In Sec. \ref{section1} we calculate the energy spectrum of the Dirac fermions, as surface as well as quantum--confined ones, confined in the cylindrical quantum wire. Besides, we find conditions, when backscattering of the surface DFs by a scalar potential is suppressed in zero magnetic field. In Sec. \ref{section2} we consider influence of the magnetic field on the spectrum of the states in the wire. It is shown that the massless surface DFs emerge periodically on the magnetic flux through the quantum wire. Finally, we explicitly determine factors that break down suppression of backscattering of the surface massless DFs in the magnetic field and calculate the conductance in this regime.

 \section{Dirac fermions in cylindrical quantum wire without magnetic field}\label{section1}

\begin{figure*}
\begin{center}
			\includegraphics[width=15cm]{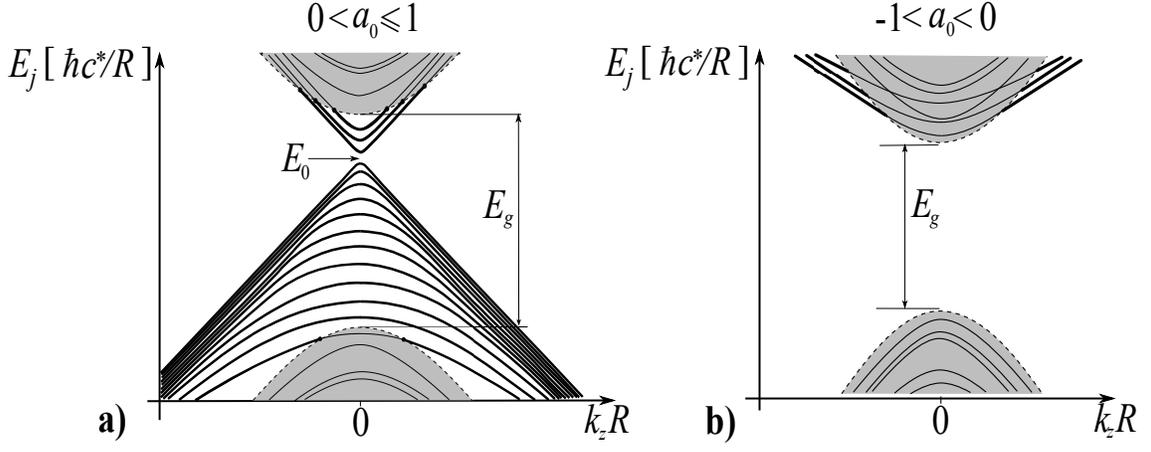}
			\caption{ \label{Fig:QW_B=0} 
 Energy spectrum $E_j(k_z)$ of states in a quantum wire without magnetic field for two classes of the surface: a) at $0<a_0\leq 1$, b) at $-1<a_0<0$; $E_g=2mc^{2}$ is the bulk gap, $R$ is a wire radius. Grey color fills regions of quantum--confined states. Solid bold lines emphasize the spectrum of the TD states, associated with different values of total angular momentum $j$. All subbands have double degeneracy due to the sign of $j$. Dash lines show boundaries between TD states and quantum--confined ones.}
\end{center}
\end{figure*}

Consider a cylindrical quantum wire with a radius $R$. Choose the cylindrical axis $z$ along the wire axis. The envelope functions of DFs $\Psi_c,\Psi_v$ obey the 3D Dirac equation (\ref{Dirac_hamiltonian}) and satisfy BCs (\ref{boundary_condition_main}) with constant $a_0$ at the cylindrical surface. From now on we set $\hbar=c=1$ everywhere, except where it is needed. Cylindrical symmetry of the 3D Dirac equation (\ref{Dirac_hamiltonian}) and BCs (\ref{boundary_condition_main}) implies conservation of longitudinal momentum $k_z$ and total angular momentum $J_z=\sigma_0\otimes j_z$, with eigenvalues $j=\pm 1/2,\pm 3/2,\dots$, where $j_z=\sigma_0(-i\partial_{\theta})+\sigma_z/2$. Hence, one can find $\Psi_c$ as follows:
\begin{equation}\label{Psi_c}
\Psi_c=\left (
\begin{array}{c}
\psi_{c1}(r)e^{i(j-1/2)\theta} \\
\psi_{c2}(r)e^{i(j+1/2)\theta}
\end{array}
\right )e^{ik_zz}.
\end{equation} 
By use of the Dirac equation (\ref{Dirac_hamiltonian}), it is convenient to express $\Psi_v$ via $\Psi_c$. The radial wave functions $\psi_{c1}(r),\psi_{c2}(r)$ obey the Bessel equation 
\begin{equation}\label{radial_DE}
\begin{array}{l}
\left (-\frac{\partial^2}{\partial r^2}-\frac{\partial }{r\partial r} + \frac{(j\mp 1/2)^2}{r^2} \right )\psi_{c1,c2}=\left (E^2-m^2-k_z^2 \right )\psi_{c1,c2},
\end{array}
\end{equation}
and BCs
\begin{equation}\label{radial_BC}
\left.
\left ( 
\begin{array}{l}
i\left ( \partial_r - \frac {j-1/2}{R_0} -a_0(E+m)\right ) \quad k_z \\
-k_z \quad   i\left ( \partial_r + \frac {j+1/2}{R_0} -a_0(E+m)\right )
\end{array}
\right )
\left (
\begin{array}{l}
\psi_{c1}(r) \\
\psi_{c2}(r)
\end{array}
\right )
\right|_{r=R}=0.
\end{equation}
Therefore $\psi_{c1,c2}(r)=C_{1,2} J_{j\mp 1/2}(kr)$, where $k=~\sqrt{E^2-m^2-k_z^2}$, $J_{j\mp 1/2}(kr)$ -- the Bessel functions of the first kind, $C_1,C_2$ -- arbitrary constants.
The BCs (\ref{radial_BC}) impose a relationship for the constants $C_1,C_2$ and give a dispersion equation
\begin{equation}\label{QW_zeroB_DE}
\begin{array}{l}
k \left [ \dfrac{J_{j-1/2}(kR)}{J_{j+1/2}(kR)} - \dfrac{J_{j+1/2}(kR)}{J_{j-1/2}(kR)} \right ]= \quad\quad\quad\quad\quad\quad\quad\\
\\
\quad\quad\quad\quad\quad\quad\quad =\left ( a_0 - \dfrac{1}{a_0} \right )E + m\left ( a_0 + \dfrac{1}{a_0} \right ). 
\end{array}
\end{equation} 
Imagine values of $k$ correspond to the TD states, real values of $k$ describe quantum--confined states of bulk DFs. Because of a symmetry property of the dispersion equation (\ref{QW_zeroB_DE}) $a_0\to 1/a_0, E\to -E$, it is enough to consider the case $|a_0|\leq 1$. Besides, the dispersion equation does not depend on $j$ sign as follows from the property of the Bessel function of integer index \mbox{$J_{-j\mp 1/2}(x)=(-1)^{j\pm 1/2} J_{j\pm 1/2}(x)$}. Consequently all 1D subbands have double degeneracy due to the $j$ sign.  Result of qualitative graphic solution of the dispersion equation (\ref{QW_zeroB_DE}) is shown on Fig.(\ref{Fig:QW_B=0}). 
The energy of quantum--confined states at $|k_z|\to~\infty$ aspires to $E=\pm\sqrt{m^2+k_z^2+\gamma_{j\pm 1/2,n}^2/R_0^2}$, where $\gamma_{j\pm 1/2,n}$ is the n-th zero of the Bessel function $J_{j\pm 1/2}$. At $a_0>0$ the TD states acquire a mass equals $v\hbar |j|/R$ in the thick wires limit ($|a_0| mcR/\hbar\gg 1$). This result is in a qualitatively agreement with the spectrum of the SSs in $Bi_2Se_3$-type quantum wires\cite{Egger_Zazunov}. The energy spectrum of the TD states has asymptotes 
\begin{equation}\label{QW_SS_Dispersion}
E=-sv\hbar\sqrt{k_z^2+\frac{j^2}{R_0^2}}+E_0,
\end{equation} 
when their decay length $\kappa^{-1}$ is much smaller than radius $R$ $(\kappa R\gg j^2) $.
The four radial components of an eigenstate $|j,k_z\rangle$ are
\begin{equation}\label{wave_function}
\begin{array}{l}
\psi_{c1}(r)=C J_{j-1/2}(kr),\\
\psi_{c2}(r)=C\dfrac{i}{k_z}\left (k - a_0(E+m)\widetilde{J} \right) J_{j+1/2}(kr),\\
\psi_{v1}(r)=C\dfrac{1}{k_z}\left ( E-m-a_0k(E+m)\widetilde{J} \right )J_{j-1/2}(kr),\\
\psi_{v2}(r)=C ia_0\widetilde{J}J_{j+1/2}(kr),
\end{array}
\end{equation}
where $\widetilde{J}=~J_{j-1/2}(kR)/J_{j+1/2}(kR)$, here and below $C$ is a normalization factor. While time reversal symmetry prohibits scattering between Kramers partners, \mbox{$(k_z,j)\to(-k_z,-j)$}, an act of backscattering, \mbox{$(k_z,j)\to(-k_z,j)$}, is allowed. 

We now show that the backscattering amplitude is suppressed in the limit $\kappa R\gg j^2 $. The wave function of the TD state (\ref{wave_function}) in this limit is
\begin{equation}\label{SS_wave_function}
|j,k_z,s\rangle\approx C
\left (
\begin{array}{c}
e^{(j-1/2)\theta} \\
-\dfrac{is}{k_z}\sqrt{k_z^2+j^2/R^2} e^{(j+1/2)\theta}\\
\dfrac{sa_0}{k_z}\sqrt{k_z^2+j^2/R^2} e^{(j-1/2)\theta}\\
ia_0 e^{(j+1/2)\theta}
\end{array}
\right )\frac{e^{\kappa(r-R)+ik_zz}}{\sqrt{2\pi\kappa r}}.
\end{equation}
Therefore the amplitude of backscattering off a scalar potential $V(\bm{r})$ is
\begin{equation}\label{matrix_element_B=0}
\langle -k_z,j,s|V|s,j,k_z\rangle\approx \frac{1}{4\pi\left ( 1+\frac{k_z^2R^2}{j^2}\right )}\int_0^{2\pi} \widetilde{ V}(R,\theta,2k_z)d\theta,
\end{equation}
where $\widetilde{V}(r,\theta,2k_z)=\int_{-\infty}^{+\infty}\exp\left ({i2k_zz} \right )V(r,\theta,z)dz/2\pi$ is the Fourier transform of $V(\bm{r})$ along $z$ axis. At integrating $\int_{0}^{R}\widetilde{ V}(r,\theta,2k_z)e^{2\kappa(r-R)}dr$, we assume that the main value of the integral comes from the vicinity of $R$. This holds true when $\widetilde{V}(r,\theta,2k_z)$ declines slower than $e^{2\kappa(r-R)}$. Otherwise the radial integral has exponential smallness in $\kappa R$, due to exponential decay of the TD wave function. From the Eq.(\ref{matrix_element_B=0}) one can see that backscattering is suppressed at $j^2/k_z^2R^2\ll 1$.

\section{Dirac fermions in cylindrical quantum wire in longitudinal magnetic field}\label{section2}

In magnetic field along the wire axis $\bm{B}=(0,0,B)$ we make the Peierls substitution $\bm{p}\to\bm{p}+e\bm{A}$ in the 3D Dirac equation (\ref{Dirac_hamiltonian}), $-e$ is the electron charge. For a vector--potential we choose the cylindrical gauge $\bm{A}=~(-By/2,Bx/2,0)$. Further, to be specific we consider the case $B>0$. The radial components $\psi_{c1},\psi_{c2}$ of the spinor $\Psi_c$ obey the following Eqs.
\begin{equation}\label{radial_DE_in_B}
\begin{array}{l}
	\left (-\dfrac{\partial^2}{\partial r^2} - \dfrac{\partial}{r\partial r} + \dfrac{(j\mp 1/2)^2}{r^2} + \dfrac{j\pm 1/2}{\lambda^2}+\dfrac{r^2}{4\lambda^4} \right )\psi_{c1,c2}= \\
	\\ 
	\qquad\qquad(E^2-m^2-k_z^2)\psi_{c1,c2}, 
\end{array}
\end{equation}
where $\lambda^2=1/eB$ is the magnetic length squared. Normalizable solutions of the radial equations (\ref{radial_DE_in_B}) are expressed in terms of the Kummer function $M(\alpha,\beta,\xi)$\cite{Abramovic}. We are interested in the TD states in the bulk gap. While $B>0$, these states have negative total angular momentum $j\leq-1/2$. Their radial components are
\begin{equation}\label{wave_function_in_B}
\begin{array}{l}
\psi_{c1}(r) = Cg(r)M\left (1-\frac{\lambda^2k^2}{2},-j+\frac 32,\frac{r^2}{2\lambda^2} \right ),\\
\\
\psi_{c2}(r)=\frac{iC}{k_zr}\left (2(j-\frac 12)- a_0R(E+m)\widetilde{M} \right )\times\\
 \qquad\qquad\qquad g(r)M\left (-\frac{\lambda^2k^2}{2},-j+\frac 12,\frac{r^2}{2\lambda^2} \right ),\\
\psi_{v1}(r)=\frac{C}{k_z}\left (E - m - \frac{a_0Rk^2\widetilde{M}}{2(j-1/2)} \right )\times\\
\qquad\qquad\qquad g(r)M\left (1-\frac{\lambda^2k^2}{2},-j+\frac 32,\frac{r^2}{2\lambda^2} \right ),\\
\psi_{v2}(r)=\frac{iCa_0R}{r}\widetilde{M}g(r)M\left (-\frac{\lambda^2k^2}{2},-j+\frac 12,\frac{r^2}{2\lambda^2} \right ),
\end{array}
\end{equation}
where \mbox{$g(r)=e^{-\frac{r^2}{4\lambda^2}}(\sqrt{2}\lambda/r)^{j-1/2}$} and
\begin{equation} \widetilde{M}=\dfrac{M(1-\lambda^2k^2/2,-j+3/2,R^2/2\lambda^2)}{M(-\lambda^2k^2/2,-j+1/2,R^2/2\lambda^2)}\nonumber.
\end{equation}

\begin{figure*}
			\includegraphics[width=16cm]{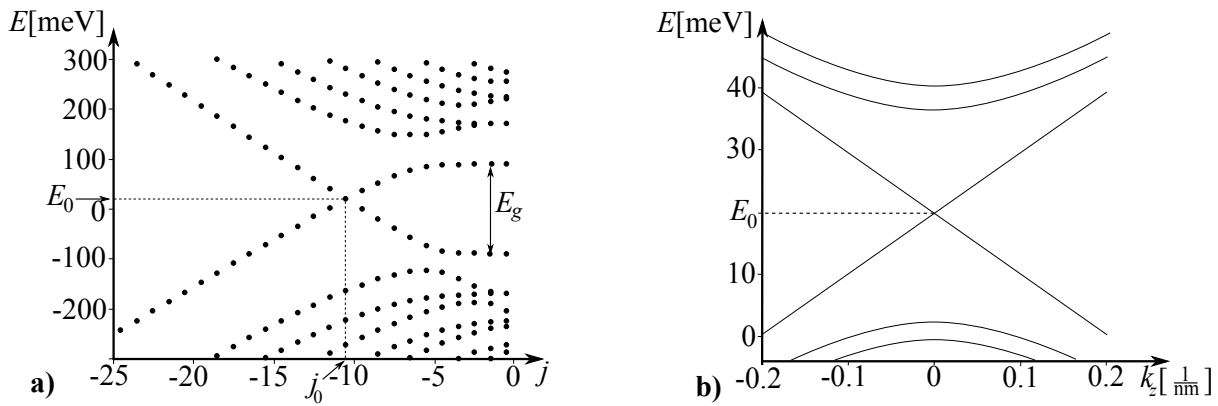}
			\caption{ \label{Fig:QW_in_B_E(j)} 
Energy spectrum of DFs in a cylinder quantum wire in a longitudinal magnetic field: a) $E_j(k_z=0)$ as a function of total angular momentum $j$; b) $E_j(k_z)$ as a function of the $k_z$ for three values of the angular momentum $j=-9.5,-10.5,-11.5$. The spectrum of the TD states with $j=j_0=-10.5$ is massless and consists from two subbands with the linear dispersion. Numerical solution was carried out at $R_0=50$nm, $a_0=0.8$, $B=6.86$T, $E_g=2mc^{2}=0.18$eV, $c=10^6$m/s. 
  }
\end{figure*}

Substituting the functions (\ref{wave_function_in_B}) in the BCs (\ref{boundary_condition_main}) we derive the dispersion equation in magnetic field for \mbox{$j\leq -1/2$}:
\begin{equation}\label{DE_in_B}
\begin{array}{l}
\left [ 2(j-1/2) - a_0R(E+m)\widetilde{M}\right ]\times\\
\quad\left [\dfrac{R^2k^2}{2(j-1/2)}+\dfrac{a_0R(E+m)}{\widetilde{M}} \right] + k_z^2R^2=0.
\end{array}
\end{equation}
Numerical solution of this equation (\ref{DE_in_B}) yields an energy spectrum $E(j,k_z)$, see Fig.\ref{Fig:QW_in_B_E(j)}.
Since in zero magnetic field the TD states is in the bulk band gap at $a_0>0$, it is reasonable to consider this class of the wire surface. A typical spectrum are shown in Fig.\ref{Fig:QW_in_B_E(j)}. In a quasiclassical limit and in a weak magnetic field ($\kappa R\gg j^2$, \mbox{$\kappa R\gg~ |j|\cdot\Phi/\Phi_0$}), the dispersion law of the TD states is 
\begin{equation}\label{QW_SS_Dispersion_in_B}
E=sv\hbar\sqrt{k_z^2+\dfrac{\left ( j+\Phi/\Phi_0\right )^2}{R^2}} + E_0,  \\
\end{equation}
where $\Phi=\pi BR^2$ is the magnetic flux through the wire cross section.

While the magnetic field is weak, the massless TD subbands periodically emerge at half-integer values of $j_0\equiv-\Phi/\Phi_0$ with a period of $\Phi_0$. With further increase of the magnetic field, the period is of weakly dependence on $B$. This effect originates from non-zero decay length of the TD states. To study properties of backscattering between the massless TD states by a scalar potential $V$, we employ the asymptotic expansion of the radial components (\ref{wave_function_in_B}) in the limit $\kappa R\gg j^2$, $\kappa R\gg~ |j|\cdot\Phi/\Phi_0$:
\begin{equation}\label{quasiqlassic_WF}
\begin{array}{l}
\psi_{c1}(r)=C \left [1-\frac{j(j-1)-2(j-\frac 32 )\frac{r^2}{2\lambda^2} }{ 2\kappa r } \right ]\left (\frac{\sqrt{2}\lambda}{r}\right )^{1/2} e^{\kappa\left (r-R_0 \right )},\\
\psi_{c2}(r)=\frac{Cs|k_z|}{ik_z} 
\left [1-\frac{j(j+1)-2(j-\frac 12)\frac{r^2}{2\lambda^2}}{2\kappa r}\right ]\left (\frac{\sqrt{2}\lambda}{r}\right )^{1/2} e^{\kappa\left (r-R_0 \right )},\\
\psi_{v1}(r)=\frac{Csa_0|k_z|}{k_z} 
\left [1-\frac{j(j-1)-2(j-\frac 32)\frac{r^2}{2\lambda^2}}{2\kappa r}\right ]\left (\frac{\sqrt{2}\lambda}{r}\right )^{1/2} e^{\kappa\left (r-R_0 \right )},\\
\psi_{v2}(r)=C ia_0\left [1-\frac{j(j+1)-2(j-\frac 12)\frac{r^2}{2\lambda^2}}{2\kappa r}\right ]\left (\frac{\sqrt{2}\lambda}{r}\right )^{1/2} e^{\kappa\left (r-R_0 \right )}.
\end{array}
\end{equation}
Calculation of the backscattering amplitude in above limit results in
\begin{equation}\label{matrix_element}
\langle j_0,-k_z,s|V|j_0,k_z,s\rangle\approx \frac{1}{2\pi}\frac{E_0}{mc^{*2}}\frac{2j_0}{\kappa R_0}\int_{0}^{2\pi}\widetilde{V}(R_0,\theta,2k_z)d\theta.  
\end{equation}
Although the amplitude does not vanish for arbitrary values of $a_0$, but it is zero if particle--antiparticle symmetry ($a_0=1$,$E_0=0$) is preserved.

To illustrate properties of backscattering processes, we calculate a conductance of the cylinder quantum wire with a length $L$ formed by the massless TD subbands in different regimes. In the collisional regime the conductance is expressed in terms of the conductivity as follows: $\Sigma=\sigma/L$. The conductivity $\sigma$ is calculated by means of the classic Boltzmann equation in the $\tau$--approximation. For simplicity we set random potential of impurity centers like this $V(\bm{r})=V_0\sum_{i}^{N}\delta(z-z_i)$, $z_i$ is a site of the $i$-th center along the wire axis, $V_0$ is power of the impurity center, $N$ is a number of the centers. Therefore, a correction $f_1(k_z)$ to the equilibrium Fermi--Dirac distribution $f_0(E)$ is 
\begin{equation}
f_1(k_z)=ev_z\tau(E)\frac{\partial f_0(E)}{\partial E}F,
\end{equation}
where $v_z=sv$, $F$ is electric field along $z$ axis, $\tau(E)$ is relaxation time. In considered approximation
\begin{equation}\label{relax_time}
\dfrac{1}{\tau(E)}=\dfrac{2L}{\hbar^2v}\langle \left | V_{-k_zk_z}\right |^2\rangle,
\end{equation}
where $\langle \left | V_{-k_zk_z}\right |^2\rangle=nV_0^2E_0^2(2j_0)^2/L(m^2c^{*4})(\kappa^2R^2)$ is the backscattering amplitude (\ref{matrix_element}) averaged over the impurity center sites, $n=N/L$ is a concentration of the impurity centers. Using of the density current formula   \mbox{$j=-e\int_{-\infty}^{+\infty}f_1(k_z)v_zdk_z/2\pi$} permits us to derive the conductance at low temperature: 
\begin{equation}\label{diffusion_conductance}
\Sigma=2\dfrac{e^2}{h}\dfrac{v\tau_F}{L},
\end{equation}
where $\tau_F\equiv~\tau(E_F)$. As the massless TD subbands emerge at half-integer values of $\Phi/\Phi_0$ in the weak magnetic fields, consequently the wire conductance will oscillate in $B$ and reach peaks at the same fluxes. However a peak magnitude is decreased with an increase of $\Phi/\Phi_0$ in this regime. It follows from the relationship $j_0\equiv-\Phi/\Phi_0$ and Eqs.(\ref{diffusion_conductance}), (\ref{relax_time}). Another effect of a magnetic field increase is deterioration of strictly $B$-periodic oscillations of the conductance. In the ballistic regime ($v\tau_F\gg L$), the conductance is obeyed the Landauer formula $\Sigma=e^2/h$. In such a case magnitude of the peak does not depend on $B$. These periodic or quasiperiodic oscillations of the conductance is a manifestation of the Aharonov--Bohm effect.  

\section{Discussion and Summary}

So, we studied the energy spectra of the massive Dirac fermions, confined in a cylindrical quantum wire. Microscopic properties of the wire surface are characterized by a real phenomenological parameter $a_0$. The latter controls the boundary conditions for the 3D Dirac equation. There are two classes of the surfaces in dependence on the $a_0$ sign. The DFs spectra in the TD states are determined by the surface class and magnetic flux $\Phi$ passing through the wire cross section. At zero magnetic field the spectrum of the TD states consists of 1D subbands indexed by the total angular momentum. For one of the surface class ($a_0>0$), there is a qualitative agreement between the TD spectra and the spectra of the topological SSs in the $Bi_2Se_3$ quantum wire\cite{Egger_Zazunov}. In the longitudinal magnetic field, the massless TD subbands emerge in the wire of the same surface class. In the limit of small decay length of the TD states ($\kappa R\gg j^2$) and in the weak magnetic field ($\kappa R\gg~ |j|\cdot\Phi/\Phi_0$), the massless TD subbands arise at half-integer values of $\Phi/\Phi_0$. In this limit, the backscattering amplitude between the massless TD states, Eq.(\ref{matrix_element}), is a product of two small parameters, $\Phi/\Phi_0\kappa R_0$ and $(1-a_0^2)$. The former defines the smallness of TD decay length, and the latter is a measure of particle-antiparticle asymmetry. At stronger magnetic field the emergence of the massless TD subbands has quasiperiodic character. 

In the case of particle--antiparticle symmetry \mbox{($a_0\to 1$)}, the backscattering between the massless TD states is suppressed and classical conductivity $\sigma$ tends to infinity as $(1-a_0)^{-2}$. It should be emphasized that the massless TD states are not protected, generally speaking, from backscattering if $a_0\neq 1$. The theory of topological SSs in the TI quantum wires \cite{Zhang_Vishwanath,Bardarson_Brouwer} does not describe this result. The $\Phi$-periodic emergence of the massless TD states can lead to the Aharonov--Bohm effect in the wire resistance. This result is in qualitative agreement with the TI-wire case.   

\begin{acknowledgments}
The authors would like to thank S.N. Artemenko for useful discussions. This work was supported by RFBR (project \#11--
02--01290).
\end{acknowledgments}

\end{document}